\documentstyle[12pt]{article}
length{\unitlength}{1in}
\begin{document}
\begin{flushright}
CERN-TH. 7289/94\\
CTP-TAMU-30/94\\
NUB-TH-3091/94
\end{flushright}
\begin{center}
{\bf SENSITIVITY OF DARK MATTER DECTECTORS TO SUSY DARK MATTER}\\
{}~\\
R. ARNOWITT$^1$\\
Center for Theoretical Physics, Department of Physics,\\
Texas A\&M University, College Station, Texas  77843-4242\\
{}~\\
PRAN NATH\footnote[2]{Permanent Address:  Department of Physics,
Northeastern University, Boston, MA  02115}\\
Theoretical Physics Division, CERN, CH-1211 Geneva 23
\end{center}
\begin{abstract}
The sensitivity of dark matter detectors to the lightest neutralino
(${\tilde {Z}_1}$) is considered within the framework of supergravity grand
unification with radiative breaking of SU(2)xU(1).
The relic density of the ${\tilde {Z}_1}$ is constrained to obey
$0.10 \leq \Omega_{\tilde {Z}_1}h^2 \leq 0.35$,
consistent with COBE data and current ~measurements of the Hubble ~constant.
Detectors can be divided
into two classes:~~those most sensitive to spin dependent
incoherent scattering of the ${\tilde {Z}_1}$
(e.g. $CaF_2$) and those most sensitive to spin independent
coherent scattering (high A nuclei e.g. Pb).
The parameter space is studied over the range of
$100GeV \leq m_0, m_{\tilde {g}} \leq 1~TeV; 2 \leq tan\beta \leq 20$;
and $-2 \leq A_t/m_0 \leq 3$ and it is found that the latter type
detector is generally more sensitive
than the former type.  Thus at a sensitivity level of
$R \geq 0.1$ events/kg da, a lead detector
could scan roughtly 30\% of the ~parameter space
studied, and an increase of ~this sensitivity by a
factor of 10 ~would lead to coverage of about 70\%
of the parameter space.  Dark matter detectors
are in general more sensitive to the high $tan\beta$,
low $m_{\tilde {g}}$ and low $m_0$ parts of the
parameter space.  The conditions of radiative
breaking of SU(2)xU(1) enter importantly in
analysing the efficiency of dark matter detectors.
 ($^1$Talk at Lake Louise Winter Institute - 1994.)\\
\end{abstract}

\noindent
{\bf 1.  Introduction}\\
{}~\\
There is strong astronomical evidence for the existance of dark matter both
within our galaxy and in other galaxies and galactic clusters.  A large number
of candidates for dark matter have been proposed both in particle physics
(neutralinos, neutrinos, sneutrinos, axions, etc.) and in astronomy (brown
dwarfs, neutron stars, black holes, Jupiters, etc.).  For supersymmetric
theories with R parity invariance, the lightest supersymmetric particle (the
LSP) is stable.  For most SUSY models and for most of the parameter space of
these models the LSP is the lightest neutralino, the ${\tilde {Z}_{1}}$.  The
dark matter (DM) for such SUSY models would then be the relic ${\tilde
{Z}_{1}}$
of the big bang which now exist in the halo of the Galaxy (and explain the
rotation curves of matter in the Galaxy).  These particles would then impinge
on DM detectors.  What we will discuss here is the ability of such detectors
to see these ${\tilde {Z}_{1}}$.\\
{}~\\
Models based purely on cold dark matter (CDM) appear to be inconsistent with
the COBE and other data, and a mix with hot dark matter (HDM) in the ratio of
$\simeq 2:1$ yields a satisfactory model.  (A candidate for HDM might be
massive neutrinos).  In addition there can be baryonic dark matter (B) at the
$\stackrel{<}{\sim}$ 10\% level.  Defining $\Omega_i = {\rho_i}/{\rho_c}$,
where $\rho_i$ is the mass density of the $i^{th}$ constituent and $\rho_c$
is the critical mass density to close the universe, then a reasonable mix is\\
{}~\\
\begin{equation}
 \Omega_{\tilde {Z}_1}\simeq 0.6; ~~\Omega_{HDM}\simeq 0.3;
{}~~\Omega_{B}\simeq 0.1
\end{equation}
{}~\\
What can be determined theoretically, however, is $\Omega_{\tilde {Z}_1}h^2$
where h = H/(100 km/s Mpc) and H is the Hubble constant.
Measurements of h give the range $h\simeq 0.5-0.75$.  Hence one has\\
{}~\\
\begin{equation}
\Omega_{\tilde {Z}_1}h^2 \cong 0.1 - 0.35
\end{equation}
{}~\\
In the following we will assume Eq. (2) is the allowed band of values for
$ \Omega_{\tilde {Z}_1}h^2$.  These bounds strongly restrict the
parameter space of
SUSY models, and hence will constrain the predictions of the SUSY DM detector
efficiencies.  Eq. (2) represents estimated bounds on
$\Omega_{\tilde {Z}_1}h^2$
and we will see that the results are a little sensitive to the lower bound.
However, lowering this bound generally will raise the detection rates, and so
Eq. (2) gives a conservative estimate of event rates.\\
{}~\\
{\bf 2.  Supergravity Gut Models}\\
{}~\\
In order to calculate $\Omega_{\tilde {Z}_1} h^2$, one needs to specify the
SUSY model.  We will use here supergravity GUT models [1].  There are a
number of advantages to this choice:
\begin{enumerate}
\item
These models are consistent with the LEP results on unification of the gauge
coupling constants at the GUT scale $M_{G} \approx 10^{16}GeV$.
\item
They generate electroweak breaking naturally by radiative corrections i.e.
using the renormalization group equations (RGE) starting at scale
$Q=M_G$ one finds a Higgs $(mass)^2$ turning negative at $Q \approx M_Z$.
\item
They depend on only four new parameters, in contrast to the low energy MSSM
which usually is chosen to have 20 new parameters (and could have as many as
137!).
\end{enumerate}
In SUSY models one needs two Higgs doublets ($H_1$ and $H_2$)
to cancel anomalies
and to give rise to masses for both u and d quarks.  Running the RGE from $M_G$
down to the electroweak scale and minimizing the Higgs potential with respect
to $\langle H_{1,2} \rangle$ one obtains two electroweak breaking equations,\\
{}~\\
\begin{equation}
{1\over 2} M_Z^2=\mu^{2}+{{\mu_{1}^2}-{\mu^{2}_2 tan^2\beta}
\over{tan^2\beta-1}};
{}~~sin2\beta = {{2m^2_3}\over{{2\mu^2}+{\mu^{2}_1}+{\mu^{2}_2}}}
\end{equation}
{}~\\
where $tan\beta\equiv\langle H_{2} \rangle/\langle H_{1}\rangle,
{\mu^{2}_i}=m_{H_i}^{2}+\Sigma_i,m_{{H}_{i}}$
are the running Higgs masses, $\Sigma_i$ are loop corrections and
$\mu$ is the running Higgs mixing parameter (which
enters in the superpotential as $-\mu H_{1} H_{2})$.  Then all SUSY masses,
widths cross sections etc., can be determined from four parameters and the sign
of $\mu$.  These may be chosen as $m_0$ (universal soft breaking spin zero
mass); $m_{\tilde {g}}$ (gluino mass); $A_t$ (t-quark cubic soft breaking
 parameter)
and $tan\beta$.  One limits the range of these parameters by imposing the
experimental lower bounds of LEP and the Tevatron on the SUSY masses, and
 also we will
require $m_{0}, m_{\tilde {g}}<1~TeV$ so that no extreme
fine tuning of parameters will occur.\\
{}~\\
Since there are 32 new SUSY particles and only 4 parameters,
 there is considerable constraint in the system.
 Two predictions that result which are relevant to
dark matter is an upper bound on the $\tilde {Z}_{1}$ mass and a lower bound
on $tan\beta$:\\
{}~\\
\begin{equation}
m_{\tilde {Z}_1} \stackrel{<}{\sim} 150 GeV;~~tan\beta > 1
\end{equation}
{}~\\
{\bf 3.  Calculation of ${\tilde {Z}_1}$ Relic Density}\\
{}~\\
For models with R parity, the ${\tilde {Z}_1}$ is absolutely stable.  However,
the primordial
${\tilde {Z}_1}$ can annihilate in the early universe,
the main diagrams being shown in Fig. 1.
 We consider the simplest approximation [2]
where at high temperature the ${\tilde {Z}_1}$ is in equilibrium with
the background.  When the annihilation rate falls below the expansion rate of
the universe, freezeout occurs at temperature $T_f$.  The ${\tilde {Z}_1}$ are
disconnected from the background and then continue to annihilate.
Thus the larger the annihilation rate, the smaller the final relic density.
The current relic density is [2]\\
\newpage
{}~\\
{}~\\
{}~\\
{}~\\
{}~\\
{}~\\
{}~\\
{}~\\
{}~\\
{}~\\
{}~\\
{}~\\
Fig. 1~~Annihilation diagrams of ${\tilde {Z}_1
{}~\\
\begin{equation}
\Omega_{\tilde {Z}_1}h^{2}\cong 2.5 \times 10^{-11}(T_{\tilde
v \rangle dx}
\end{equation}
{}~\\
Here ${\langle~~~\rangle}$ means thermal average and $\sigma$ is the
annihilation
cross section at relative velocity $v$.  Freezeout generally occurs when the
${\tilde {Z}_1}$ is non-relativistic i.e. ${x_{f}}\equiv kT_{f}/m_{\tilde
{Z}_1}\simeq {1\over20}$, and so the thermal average can be taken using the
Boltzman distribution:\\
{}~\\
\begin{equation}
\langle \sigma v \rangle = {\int_0^\alpha} dv {v^2} (\sigma v)
exp[{-v^2}/4x]/{\int_0^\alpha} dv {v^2} exp[-{v^2}/4x]
\end{equation}
{}~\\
This has led in the past to using a non-relativistic expansion for $\sigma v$
i.e. $\sigma v\cong a + b(v^{2}/c^{2})+$..., with which it becomes trivial to
take the thermal average.  However, as has been pointed out [3], this
 expansion can be
a bad approximation near a narrow s-channel pole, even though
$v^{2}/c^{2} <<1$.  For the ${\tilde {Z}_1}$,
the approximation turns out generally
to be quite bad near the Higgs and $Z^0$ pole as indicated in Fig. 2 [4].\\
{}~\\
Note that the non-relativistic expansion for $\sigma v$ can fail over
a wide range of $m_{\tilde {g}}$ (due to the smearing of the thermal
 averaging and often closeness of the h and Z poles.).\\~\\
{}~\\
{}~\\
{}~\\
{}~\\
{}~\\
{}~\\
{}~\\
{}~\\
{}~\\
{}~\\
{}~\\
{}~\\
{}~\\
{}~\\
{}~\\
{}~\\
Fig. 2~~$\Omega$approx/$\Omega$ vs $m_{\tilde{g}}$ for $m_0=700$ GeV,
$tan\beta=2.25, A_t=0$, $\mu>0$ and $m_t=140$ GeV.
$\Omega$approx is calculated using $\sigma v=a+bv^2$ while $\Omega$ is
calculated by accurate evaluation of the
integrals of Eqs. (6,7) numerically.  The h and Z poles are where the curves
go from positive to negative values.\\
{}~\\
In general, the allowed values of $\Omega_{\tilde {Z}_1}h^{2}$ of Eq.(2)
are quite
small, implying the need for a large amount of annihilation to occur.  This
happens in the regions of parameter space where $2m_{\tilde {Z}_1}$ is close to
$m_h$ or $M_Z$ or when the slepton/squark masses are small (enhancing the
t-channel poles).  The former occurs commonly in many models, while the latter
can occur when $m_0$ is small, as is the case in the no-scale models [4] (where
$m_0$ is zero).\\
{}~\\

{\bf 5.  Expected Detector Event Rates}\\
{}~\\
In the supergravity Gut models, the value
of R depends on the point in thi
tion}
100GeV \leq m_0, m_{\tilde {g}} \leq 1TeV;~~ -2m_0 \leq A_t \leq 3.5m_0;
{}~~ 2 \leq tan\beta \leq 20
\end{equation}
{}~\\
The parameter space is further restricted by the requirements that
$(i) 0.10 \leq \Omega_{\tilde {Z}_1}h^2 \leq 0.35$, (ii)
experimental bounds on SUSY masses from LEP and the Tevatron are not
violated, and (iii) radiative breaking [Eq. (3)] of SU(2)xU(1) occurs.
Event rates for the following detectors have been analyzed:\\
{}~\\
\begin{equation}
^{3}He;~~^{40}Ca^{19}F_{2};~~^{76}Ge + ^{73}Ge;~~^{71}Ga ^{75}As;
{}~~^{23}Na ^{127}I;~~^{207}Pb
\end{equation}
{}~\\
The first two detectors have large spin dependent ${\tilde {Z}_1}$ scattering
($CaF_2$ being the largest) while
the last four have increasingly large spin independent coherent
${\tilde {Z}_1}$ scattering.  We will see that
throughout most of the parameter space the heavy nuclei with large coherent
scattering are more efficient
detectors than those with the large spin dependent scattering.\\
{}~\\
While the different parameters of Eq. (17) which define the theory enter
 in many different places in the calculation
(e.g. in calculating the relic density in Sec. 2, the ${\tilde {Z}_1}$
coefficients of Eq. (11), $A_q$, $B_q$
and $C_q$ of Eqs. (10), (13) etc.) it is possible to exhibit the general
dependence of the event rate on them.\\
{}~\\
{}~\\
{}~\\
{}~\\
{}~\\
{}~\\
{}~\\
{}~\\
{}~\\
{}~\\
{}~\\
{}~\\
{}~\\
{}~\\
{}~\\
{}~\\
{}~\\
{}~\\
{}~\\
\begin{center}
\begin{Large}
$m_{\tilde {g}}$ [GeV]
\end{Large}
\end{center}
{}~\\
Fig. 4~~R[Pb] (solid) and R [$CaF_2$] (dashed) vs $m_{\tilde {g}}$ for
$tan\beta$=6 (lower curves) and
$tan\beta$=20 (upper curves).  $A_t$=1.5 $m_0$, $m_0$=100GeV, and $\mu>0$.\\
{}~\\
Fig. 4 shows that the event rate decreases with the gluino mass, and that
the Pb detector (largest coherent
scattering) has considerably higher event rates than $CaF_2$ (largest
spin dependent scattering).  The decrease
of R with $m_{\tilde {g}}$ arises from the fact that as $m_{\tilde {g}}$
 increases, the ${\tilde {Z}_1}$ becomes more and more
Bino, and one needs an interference between the Bino and Higgsino
parts of ${\tilde {Z}_1}$ to generate sizable
scattering in Eq. (13).  Note also that the $tan\beta$=20
curve lies higher then the $tan\beta$=6 curve.  This
is in part due to the $1/{cos\beta}$ factor in the $C_d$
contribution of Eq. (13) yielding a $tan^{2}\beta$
dependence for the $C_d$ part of the coherent scattering.
  This behavior can be seen more explicitly in Fig. 5.
At fixed $tan\beta, m_0$ has been chosen in Fig. 5 so
that $\Omega_{\tilde {Z}_1}h^2$ is approximately equal for each
curve.  Thus Fig. 5 also shows that the event rate increases
 with $A_t$ for a fixed value of $\Omega_{\tilde {Z}_1}h^2$.
The NaI curves lie higher than the Ge curves since $^{127}I$
 has higher nuclear mass than $^{76}Ge + ^{73}Ge$.\\
\newpage
{}~\\
{}~\\
{}~\\
{}~\\
{}~\\
{}~\\
{}~\\
{}~\\
{}~\\
{}~\\
{}~\\
{}~\\
\begin{center}
\begin{Large}
$tan\beta$\\
\end{Large}
\end{center}
{}~\\
Fig. 5~~R vs $tan\beta$ for NaI and Ge detectors for
 $m_{\tilde {g}}=275GeV, \mu > 0$.  The solid
curve is for $A_t/m_0=0.0$, $m_{0}=200GeV$, dashed curve for
$A_t/m_0=0.5$, $m_0=300GeV$, and dash-dot
curve for $A_t/m_0=1.0$, $m_0=200GeV$.
For each pair the upper curve is for NaI and the lower for Ge.\\
{}~\\
{}~\\
{}~\\
{}~\\
{}~\\
{}~\\
{}~\\
{}~\\
{}~\\
{}~\\
{}~\\
{}~\\
{}~\\
{}~\\
{}~\\
\begin{center}
\begin{Large}
$m_0$[GeV]\\
\end{Large}
\end{center}
{}~\\
\noindent
Fig. 6~~R vs $m_0$ for $m_{\tilde {g}} = 300GeV, A_t/m_0=0.5,
tan\beta=8, \mu >0$.  The dashed curve is
for $CaF_2$, the solid curves (from bottom to top) are for Ge, NaI and Pb.\\
\newpage
Fig. 6 shows that R decreases with $m_0$, as one might expect since
the slepton pole contribution decreases.
(Actually, the situation is more complicated as $m_0$ enters
in the radiative breaking equations, Eq. (3), which
determine $\mu$, and $\mu$ enters in the nuetralino
mass matrix which determines $\alpha, \beta, \gamma, \delta$
of Eq. (11)).  For the particular parameters chosen in Fig. 6,
 the incoherent spin-dependent scattering
for the $CaF_2$ detector exceeds the coherent s
cattering seen by the other detectors.  (This type
situation rarely occurs).  Note also that the Ge,
NaI and Pb curves sequence themselves in the order of their
atomic numbers.\\
{}~\\
Fig. 7 shows the maximum and minimum event rates for the $CaF_2$
and Pb detectors as a function of $A_t$ as one
varies all other parameters over the entire parameter\\
{}~\\
{}~\\
{}~\\
{}~\\
{}~\\
{}~\\
{}~\\
{}~\\
{}~\\
{}~\\
{}~\\
{}~\\
{}~\\
{}~\\
{}~\\
{}~\\
{}~\\
{}~\\
{}~\\
{}~\\
Fig. 7~~Maximum and minimum event rates of $CaF_2$
(dashed curve) and Pb (solid curve) detectors for $\mu>0$
as one varies all parameters.\\
{}~\\
space:~~$2 \leq tan\beta \leq 20; 100GeV \leq m_0, m_{\tilde {g}} \leq 1TeV;
-2 \leq A_t/m_0 \leq 3.5$.
One sees that the Pb detector exceeds the $CaF_2$ detector in sensitivity
by a factor of 5-10.  The very
large event rates all come from the largest $tan\beta$,
while the minimum rates come from different smaller
values of $tan\beta$ for different $A_t$.  The expected rates
for the other detectos (e.g. Ge, NaI etc.)
scale approximately with the Pb detector by their atomic numbers.\\
\newpage
\noindent

{\bf 6.  Conclusions}\\
{}~\\
We have examined here the expected sensitivity of dark matter detectors to
udes the constraint of radiative breaking of SU(2)$\times$U(1)
as well as the requirement that
$0.10 \leq \Omega_{\tilde {Z}_1}h^2 \leq 0.35$
consistent with the COBE data.  Detectors fall
into two catagories:  those that are sensitive
to the spin dependent incoherent ${\tilde {Z}_1}$
scattering (e.g. $CaF_2$) and those most senstive
to the coherent scattering (e.g. Pb).  The latter
 have event rates that increase with the nuclear mass, and
hence favor heavy nuclei.  We find in general that throughout
almost all the parameter space, the latter type
detectors are significantly more efficient than the former.\\
{}~\\
In general the dark matter detectors are more sensitive to
the high $tan\beta$ and small $m_{\tilde {g}}$ part of the
parameter space (and generally the small $m_0$ part).  At
the current expected sensitivity of $R>0.1$ events/kg da,
one can expect to examine about 20-30\% of the parameter
space (using high A nuclei) for $\mu>0$. (The $\mu<0$ event
rates are generally smaller.)
An increase in the sensitivity by a factor of 10 would enable
these detectors to examine 60-70\% of the parameter space.
One would need a sensitivity of $R>0.001$ events/kg da to cover
the entire parameter space.\\
{}~\\
Since the assumption that the ${\tilde {Z}_1}$ is the cold dark matter is high
r dark matter in part of the parameter space will be a significant aid
in reducing the ambiguities that
exist in supergravity Gut models.\\
{}~\\
{\bf Acknowledgements}\\
{}~\\
This research was supported in part by NSF Grant Nos. PHY-916593 and
PHY-917809.\\
{}~\\
{\bf References}\\
\begin{enumerate}
\item
For reviews see P. Nath, R. Arnowitt and A.H. Chamseddine, "Applied N=1
Supergravity, {\it ICTP Lecture Series}~{\bf  Vol. 1} 1983 (World Scientific,
Singapore, 1984); H.P. Nilles, {\it Phys. Rep.} {\bf 110}, 1 (1984).
\item
B.W. Lee and S. Weinberg, {\it Phys. Rev. Lett.} {\bf 39}, 165(1977);
D.A. Dicus, E. Kolb and V. Teplitz, {\it Phys. Rev. Lett.} {\bf 39}, 168
(1977); H. Goldberg, Phys. {\it Rev. Lett.} {\bf 50}, 1419 (1983);
J. Ellis, J.S. Hagelin, D.V. Nanopoulos, K. Olive and N. Srednicki,
{\it Nucl. Phys.} {\bf B238}, 453 (1984).
\item
K. Greist and D. Seckel, {\it Phys. Rev.}{\bf D43}, 3191 (1991);
P. Gondolo and G. Gelmini,{\it  Nucl. Phys.} {\bf B360}, 145 (1991).
\item
R. Arnowitt and P. Nath, {\it Phys. Lett.} {\bf B299}, 58, (1993);
{\bf B303}, 403 (1993)(E); P. Nath and R. Arnowitt, {\it Phys. Rev.
Lett.} {\bf 70}, 3696 (1993).
\item
M.W. Goodman and E.Witten, {\it Phys. Rev.} {\bf D31}, 3059 (1985);
K. Greist, {\it Phys. Rev.} {\bf D38}, 2357 (1988); {\bf D39}, 3802 (1989)((E);
R. Barbieri, M. Frigeni and G.F. Giudice, {\it Nucl. Phys.} {\bf B313},
725 (1989); M. Srednicki and R. Watkins, {\it Phys. Lett.}{\bf  B225}, 140
(1989);
J. Ellis and R. Flores, {\it Phys. Lett}. {\bf B265}, 259 (1991);
{\bf B300} 175 (1993); {\it Nucl. Phys.} {\bf B400}, 25 (1993);
M. Kamionkowski, {\it Phys. Rev.} {\bf D44}, 3021 (1991);
M. Drees and M. Nojiri, {\it Phys. Rev.}{\bf D48}, 3483 (1993).
\item
H. Haber and R. Hempfling, {\it Phys. Rev. Lett.} {\bf 66}, 1815 (1991);
{\it Phys. Rev.} {\bf D48}, 4280 (1993);
J. Ellis, G. Ridolfi and F. Zwirner, {\it Phys. Lett.} {\bf B262}, 477 (1991).
\end{enumerate}
\end{document}